\documentclass[
aps,
amsmath,amssymb,
prl,
reprint,
groupedaddress,
superscriptaddress,
]{revtex4-2}

\usepackage{graphicx}
\usepackage{dcolumn}
\usepackage{bbm}
\usepackage{bm}
\usepackage{hyperref}
\hypersetup{colorlinks=true,citecolor=blue,urlcolor=blue,linkcolor=blue}
\usepackage{xcolor}

\begin{document}


\title{Photothermal Fourier-plane Phase Synchronization for Interferometric Scattering Microscopy}

\author{Shupei Lin}
\altaffiliation{These authors contributed equally to this work.}
\affiliation{School of Physics and Wuhan National Laboratory for Optoelectronics, Institute for quantum science and engineering, Huazhong University of Science and Technology, Luoyu Road 1037, Wuhan, 430074, People's Republic of China}

\author{Nanfang Jiao}
\altaffiliation{These authors contributed equally to this work.}
\affiliation{School of Physics and Wuhan National Laboratory for Optoelectronics, Institute for quantum science and engineering, Huazhong University of Science and Technology, Luoyu Road 1037, Wuhan, 430074, People's Republic of China}

\author{Yevhenii Shaidiuk}
\affiliation{Institute of Photonics and Electronics of the Czech Academy of Sciences, Chabersk{\'a} 1014/57, 18251 Prague, Czech Republic}

\author{Delong Feng}
\affiliation{School of Physics and Wuhan National Laboratory for Optoelectronics, Institute for quantum science and engineering, Huazhong University of Science and Technology, Luoyu Road 1037, Wuhan, 430074, People's Republic of China}

\author{Jingwei Luo}
\affiliation{School of Physics and Wuhan National Laboratory for Optoelectronics, Institute for quantum science and engineering, Huazhong University of Science and Technology, Luoyu Road 1037, Wuhan, 430074, People's Republic of China}

\author{Yihao Yu}
\affiliation{School of Physics and Wuhan National Laboratory for Optoelectronics, Institute for quantum science and engineering, Huazhong University of Science and Technology, Luoyu Road 1037, Wuhan, 430074, People's Republic of China}

\author{{\L}ukasz Bujak}
\affiliation{Institute of Photonics and Electronics of the Czech Academy of Sciences, Chabersk{\'a} 1014/57, 18251 Prague, Czech Republic}

\author{Jianwei Tang}
\affiliation{School of Physics and Wuhan National Laboratory for Optoelectronics, Institute for quantum science and engineering, Huazhong University of Science and Technology, Luoyu Road 1037, Wuhan, 430074, People's Republic of China}

\author{Marek Piliarik}
\affiliation{Institute of Photonics and Electronics of the Czech Academy of Sciences, Chabersk{\'a} 1014/57, 18251 Prague, Czech Republic}

\author{Xue-Wen Chen}
\email[To whom correspondence should be addressed:\\ ]{xuewen\_chen@hust.edu.cn}
\affiliation{School of Physics and Wuhan National Laboratory for Optoelectronics, Institute for quantum science and engineering, Huazhong University of Science and Technology, Luoyu Road 1037, Wuhan, 430074, People's Republic of China}
\affiliation{Wuhan Institute of Quantum Technology, Wuhan 430206, People's Republic of China}

\begin{abstract}
We introduce and experimentally implement Fourier-plane phase synchronization for optical microscopy, and demonstrate its performance with interferometric scattering microscopy. By combining a photothermal phase plate and laser beam scanning, we realize a synchronized phase for all scattering components on the Fourier plane of high numerical-aperture microscopes, where the evanescent waves and optical aberration normally produce highly inhomogeneous phase distributions. We achieve an almost perfect point spread function, exhibiting a tighter focus with 50\% enhancement of the signal and ideal circular symmetry. Particularly, by synchronizing the phase to $\pi/2$, we demonstrate the background speckles exhibit an anti-symmetric dependence on axial defocus, enabling the effective suppression of the speckles via defocus integration and thus the detection of 10 nm particles immobilized on the substrate. The concept and technique of seamless dynamic phase control on the Fourier plane constitute a key asset for modern optical microscopy.
\end{abstract}

\maketitle

Since its inception, optical microscopy has been a cornerstone of scientific discovery, enabling noninvasive visualization of structures and dynamics at the microscopic scale \cite{Mertz2019IntrodOptMicroscopy}. At the heart of every imaging modality lies the point spread function (PSF), which characterizes the system’s response to a point object and forms the basis for image interpretation and quantitative analysis \cite{Cole2011NatProt,Weisenburger2015ContempPhys}. The formation of the PSF becomes particularly critical for advanced quantitative imaging techniques \cite{Zhou2019APLPhotonics,Park2018NatPhotonics,Shechtman2014PRL,Xie2014PRL} as its shape and symmetry directly affect measurement fidelity \cite{KIRSHNER2013JMicroscopy,Parthasarathy2012NatMethod,Diezmann2017ChemRev,Pushkina2021PRL}. However, the clarity and symmetry of the PSF are often compromised by angle(wavevector)-dependent inhomogeneous phase responses, background scattering, and optical aberrations \cite{Yu2010PRL,Gjonaj2013PRL}. These issues are especially severe in interferometric techniques\textemdash including interferometric scattering (iSCAT) microscopy \cite{Ginsberg2025NatRevMethodsPrim}, holographic microscopy \cite{Kim2021DigHoloMicro}, and interference reflection microscopy \cite{Barr2009CurrentProtCellBio}\textemdash which rely on precise phase differences between the scattered and reference fields. Although our experiments focus on iSCAT, the effects discussed here are generic to all brightfield interferometric modalities. 

iSCAT could provide ultrahigh sensitivity with the utilization of effective background suppression, and has enabled a broad range of applications in fundamental science and technology. It has been employed for tracking single nanoparticles \cite{Huang2017ACSNano,Kashkanova2022NatMethod,Huang2024LSA}, detecting single proteins \cite{Piliarik2014NatComm,Ortega2014NanoLett} and their masses \cite{Young2018Science}, probing the dynamics of biological membranes and ion channels \cite{Taylor2019NatPhotonics,Li2025NatPhotonics}, visualizing chemical reactions \cite{Gruber2024Nature}, and monitoring carrier migration in optoelectronic materials \cite{Delor2020NatMater,Merryweather2021Nature}\textemdash achievements largely facilitated by dynamic background subtraction that isolates the signal of interest. However, even in such cases, the PSF often exhibits distortions and asymmetries, especially in high-numerical-aperture (NA) systems where evanescent components possess distinct phase responses from propagating ones at the interfaces \cite{Novotny2012PrinNanoOpt} and where optical aberration are pronounced \cite{Eismann2021LSA}. The challenge becomes more acute for stationary specimens, where speckle backgrounds arising from substrate roughness \cite{Lin2022PRL} cannot be removed via dynamic subtraction. As a result, direct wide-field optical detection of immobilized nanoparticles below 20 nm remains difficult for iSCAT  \cite{Failla2006NanoLett,Avci2017Optica,Wang2021AnalChem} and other wide-field optical microscopy techniques \cite{Meng2021AcsPhotonics,Cho2018OL,Sofronov2021PRAppl}. Yet, such static detection is of great interest, for instance, for high-throughput semiconductor wafer inspection \cite{Zhu2022IJEM,Reinhardt2018HBSiWaferCleanTech}, contamination detection in photonic circuits \cite{Brandt1973AO,Su2022EquipMaterClean}, and biomolecular assays on functionalized templates \cite{Bujak2021SmallMethods}. 

\begin{figure}
	\centering
	\includegraphics{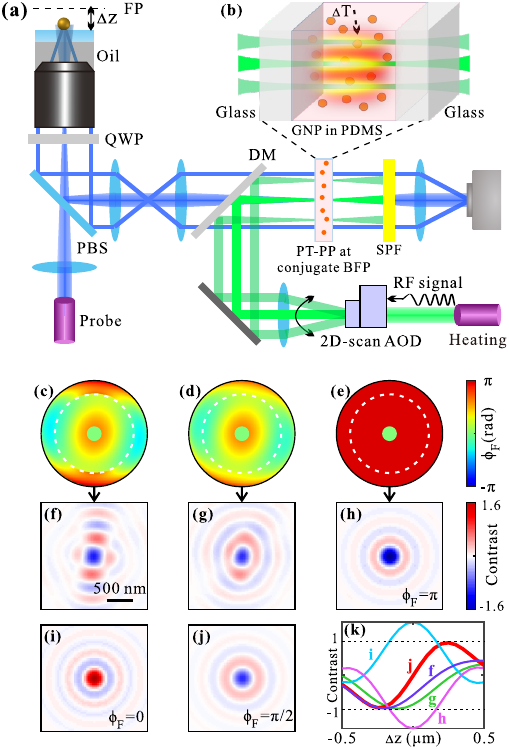}
	\caption{
		(a) Schematic of the experimental setup. (b) Structure of the photothermal-phase plate (PT-PP) and the illustration of the light-absorption induced temperature change $\Delta T$. Calculated phase profiles on the Fourier plane ($\phi _ F$) for a typical iSCAT system with an aberration level of $0.4\pi$ under (c) linearly and (d) circularly polarized illuminations. (e) Phase profile with all scattering components synchronized ($\phi _ F = \pi$). The white-dashed and central green circles in (c), (d) and (e) denote NA = 1.0 and the reference, respectively. (f), (g) and (h) Calculated iSCAT PSFs based on the phase profiles in (c), (d) and (e), respectively.  (i) and (j) are the PSFs with  $\phi _ F = 0$ and $\pi/2$, respectively. $\Delta z$ in (f)$\sim$(j) are chosen to obtain the maximum contrast. (k) Contrast evolutions with the defocus for (f)$\sim$(j). PBS: polarized beam splitter, QWP: quarter-wave plate, DM: dichroic mirror, BFP: back focal plane, PDMS: polydimethylsiloxane, SPF: short pass filter, AOD: acoustic optical deflector, RF: radio frequency.
	}
	\label{FIG1}
\end{figure}

Inspired by recent advances in adaptive optics \cite{Hampson2021NatRevMethodsPrim}, we here introduce and experimentally demonstrate the concept of phase synchronization for all scattering components on the Fourier plane to enhance iSCAT. In widefield iSCAT, the scattering components occupy the entire Fourier plane, while the reference appears as a single bright spot. Through phase synchronization, all Fourier components of the scattering—both within and beyond the critical angle in high-numerical-aperture microscopes—are aligned to share a uniform phase difference with the reference beam. Instead of conventional spatial phase modulators \cite{Hampson2021NatRevMethodsPrim,Yang2023OptElectScience,Roggemann1997RevModPhys} based on pixelated or segmented elements, we employ a photothermal phase plate (PT-PP) driven by a scanning heating laser to achieve seamless, on-demand phase synchronization. We demonstrate that the iSCAT PSF can be optimized to possess near-perfect circular symmetry, tighter focus with 50\% enhancement of the interference contrast, and effective suppression of speckle background. The resulting enhancement permits robust detection of immobilized nanoparticles as small as 10 nm that would otherwise remain obscured by speckles.  


iSCAT images the interference between a reference field $\bm{E}_{\rm r}$ and scattering field  $\bm{E}_{\rm s}$. For $\left| \bm{E}_{\rm r} \right| \gg \left| \bm{E}_{\rm s} \right|$, the interference contrast can be given as 
\begin{equation}
	c \approx 2\left| \bm{E}_{\rm s} / \bm{E}_{\rm r} \right| \cos \Delta \Phi
	\label{Eq1}
\end{equation}
where $\Delta \Phi$ is the phase difference between the two fields. In a typical wide-field configuration, the reference field can be approximated as a planewave at normal incidence.  $\bm{E}_{\rm s}$ is linked to the electric field in Fourier plane $\bm{E}_{\rm F}$ according to
\begin{equation}
	\bm{E}_{\rm s}(x,y) = \frac{if_0}{k_0} \int\!\bm{E}_{\rm F}(k_x,k_y,\Delta z)e^{i(k_x x+k_y y)}\, {\rm d}k_x{\rm d}k_y
	\label{Eq2}
\end{equation}
where $k_x=k_0 x_{\rm b}/f_0$ and $k_y=k_0 y_{\rm b}/f_0$ are the normalized Fourier components with $f_0$, $k_0$ and $(x_{\rm b},y_{\rm b})$ being the focal length of the imaging lens, wavenumber in vacuum, and the coordinate at the conjugated back focal plane (BFP) of the microscope system,respectively. As shown in FIG. \ref{FIG1}(a), $\Delta z$ is the defocus, i.e., the distance from the focus. $\bm{E}_{\rm F}$ can be expressed as
\begin{equation}
	\bm{E}_{\rm F}(k_x,k_y,\Delta z) = A_{\rm F}(k_x,k_y)e^{i[\phi_{\rm F}(k_x,k_y)+nk_0\Delta z \cos \theta]}
	\label{Eq3}
\end{equation}
where $A_{\rm F}(k_x,k_y)$ and $\phi_{\rm F}(k_x,k_y)$ are the wavevector-dependent amplitude and phase of $\bm{E}_{\rm F}$ for $\Delta z=0$, respectively. The second term $nk_0\Delta z \cos \theta$ is the phase acquired when there is a nonzero $\Delta z$ with $n$ being the refractive index of media between the sample and objective (e.g., immersion oil) \cite{He2021JPD}. Here $\theta$ is related to the Fourier component through $n \sin \theta =M\sqrt{(k_x^2+k_y^2 )}/k_0 $ with $M$ being the magnification of the imaging system. For clarity, let us consider the situation where the image center is at $x = 0$ and $y = 0$. $\left| \bm{E}_{\rm F} \right|$ reaches the maximum, when $\phi_{\rm F}(k_x,k_y)$ for all $(k_x,k_y )$, including $M\sqrt{(k_x^2+k_y^2 )}/k_0>1$, is synchronized to a constant $\phi_{\rm F}$. The iSCAT PSF can, in fact, be manipulated by adjusting $\phi_{\rm F}$. For instance, $\phi_{\rm F}=0$ or $\pi$ provides the maximum interference peak or dip with the tightest PSF achievable for a given optical imaging system. Of particular interest here is the case of $\phi_{\rm F}=\pi/2$, which makes that the interference contrast is anti-symmetric with respect to the defocus, i.e., $c(\Delta z)=-c(-\Delta z)$. This arises because the phase contribution  $nk_0\Delta z \cos \theta$ is an odd function of $\Delta z$, yielding 
$\Delta \Phi(\Delta z)+\Delta \Phi(-\Delta z)=\pi$ for all $\theta$. The property can be utilized to suppress the speckle background arising from sub-nanometer surface undulations \cite{Lin2022PRL,Jiao2024OL}.

In practice, $\phi_{\rm F}(k_x,k_y)$ is highly inhomogeneous in high-NA imaging systems and in systems subject to aberration. FIG. \ref{FIG1}(c) displays $\phi_{\rm F}(k_x,k_y)$ for a typical iSCAT system with NA = 1.35 and an aberration level of $0.4\pi$ $(0.2\lambda)$, illustrating severe phase nonuniformity. The central small green spot denotes the phase of the reference. The white-dashed circle indicates NA = 1.0, corresponding to the critical angle of the air-glass interface. Beyond NA = 1.0, $\phi_{\rm F}(k_x,k_y)$ is distinct from those within the dashed circle due to the abrupt change of the transmission coefficient from the propagating to the evanescent Fourier components. As shown in FIG. \ref{FIG1}(d), the magnitude of this phase nonuniformity decreases partly when the illumination polarization is changed from linear to circular. The situation with the scattering phase synchronized to $\phi_{\rm F}(k_x,k_y)=\pi$ for the circular-polarization illumination is depicted in FIG. \ref{FIG1}(e). FIG. \ref{FIG1}(f)$\sim$\ref{FIG1}(h) display the calculated real-space iSCAT PSFs corresponding to FIG. \ref{FIG1}(c)$\sim$\ref{FIG1}(e), respectively. Note that here in each case the defocus $\Delta z$ is adjusted to generate the maximum interference contrast. One sees that from left to right, the PSFs become tighter and more circularly symmetric. Comparing FIG. \ref{FIG1}(f) and \ref{FIG1}(h), the interference contrast is increased by 61\%. FIG. \ref{FIG1}(i) and \ref{FIG1}(j) show the calculated real-space iSCAT PSFs with the phase synchronized to 0 and $\pi/2$, respectively. As expected, both the iSCAT PSFs have perfect circular symmetry, and the case of $\phi_{\rm F}(k_x,k_y)=0$ provides the maximum interference peak. A more symmetric and sharper PSF is favorable for tracking single nanoparticles with high precision. The evolutions of the interference contrast over the defocus are illustrated by the color-coded traces in FIG. \ref{FIG1}(k). Intriguingly, for the case of $\phi_{\rm F}(k_x,k_y)=\pi/2$ (the red curve denoted by \textbf{j}), the interference contrast exhibits a perfect anti-symmetric dependence on the defocus.

As shown in FIG. \ref{FIG1}(a) and \ref{FIG1}(b), to implement phase synchronization, we employ a PT-PP and a heating laser ($\lambda$ = 520 nm) in combination with two orthogonal acoustic optical deflectors (AODs). As depicted in FIG. \ref{FIG1}(b), the PT-PP consists of two glass plates sandwiching a 1-mm-thick polydimethylsiloxane (PDMS) layer embedded with 50 nm gold nanoparticles (GNPs) \cite{Robert2022AOM,SM}. The GNPs absorb light from the focused heating laser and increase the local temperature of the PDMS. This leads to a decrease in the local refractive index, causing a change to the phase of the iSCAT probe beam. As shown in FIG. \ref{FIG1}(a), a periodic two-dimensional (2D) scan of the heating beam on the conjugated BFP (i.e., Fourier plane) allows an on-demand profile of the phase change. With the power of the heating laser fixed, its dwell-time on each position controls the phase change. 

\begin{figure}
	\centering
	\includegraphics{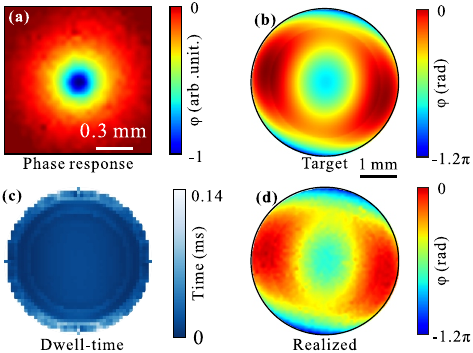}
	\caption{(a) Measured normalized profile of the photothermal phase change due to a single heating laser spot. (b) Target profile of the phase change for achieving phase synchronization  $\phi_F=\pi/2$ for a point scatterer on the coverslip surface. (c) The dwell-time distribution of the heating laser. (d) Measured profile of the photothermal phase change.
	}
	\label{FIG2}
\end{figure}

The heating laser has a spot size of 0.18 mm (FIG. S5 \cite{SM}) and induces phase changes over a finite area on the PT-PP. As depicted in FIG. \ref{FIG2}(a), we experimentally measured such ``PSF'' by applying an interferometric technique \cite{Takeda1982JOSA}. The ``PSF'' shows a much larger range than the laser spot due to balanced heat conductance and dissipation. For a prescribed laser pattern, the phase-change profile is the convolution of the laser pattern with the phase-change ``PSF''. In our experiment, the conjugated BFP has a diameter of 4.05 mm. We use a circular region of the PT-PP with a diameter of 4.7 mm enclosing the conjugated BFP and define a pixel size of 0.1 mm$\times$0.1 mm. We first obtain the target profile of the phase change as shown in FIG. \ref{FIG2}(b) to realize a phase synchronization of $\phi_F=\pi/2$ for a point scatterer lying on the substrate surface. By applying an iterative process (Supplementary Section S5 \cite{SM}) and considering the deflection efficiency of the AODs (FIG. S6 \cite{SM}), we obtain the dwell time of the heating laser for each pixel as displayed in FIG. \ref{FIG2}(c). One complete 2D scan takes 36 ms, which is less than 1/30 of the response time of the PT-PP (FIG. S7 \cite{SM}) and thus ensures a quasi-static phase change pattern (Supplementary Section S4.3 \cite{SM}). FIG. \ref{FIG2}(d) shows the measured profile of the phase change, consistent with the target profile (FIG. \ref{FIG2}(b)).

\begin{figure}
	\centering
	\includegraphics{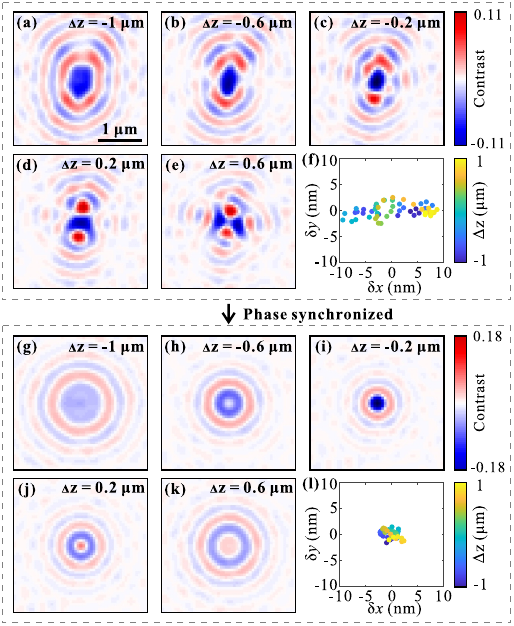}
	\caption{Measured iSCAT contrast images for a 40 nm GNP on a glass coverslip when $\Delta z$ is (a) -1 $ \rm{\mu m}$, (b) -0.6 $\rm{\mu m}$, (c) -0.2 $\rm{\mu m}$, (d) 0.2 $\rm{\mu m}$, and (e) 0.6 $\rm{\mu m}$ with no phase synchronization. (f) Position localization uncertainties of the GNP at varied defocus in the range of -1 $\rm{\mu m}$ to 1 $\rm{\mu m}$. (g)$\sim$(l) represent the same as in (a)$\sim$(f) with phase synchronization. 
	}
	\label{FIG3}
\end{figure}

For iSCAT imaging, we apply circular-polarization illumination ($\lambda$ = 490 nm) with a polarization beam splitter (PBS) and a quarter-wave plate (QWP). The scattered light and the reflected illumination (as reference) are collected through an oil-immersion objective (NA = 1.35) and pass through the conjugated BFP. FIG. \ref{FIG3}(a)$\sim$\ref{FIG3}(e) display the iSCAT contrast images of a 40 nm GNP without the phase synchronization at defocus positions of $\Delta z$ = -1.0 $\mu$m, -0.6 $\mu$m, -0.2 $\mu$m, 0.2 $\mu$m, and 0.6 $\mu$m, respectively. These images exhibit distorted patterns with the evolution of the defocus and poor circular symmetry in general. Phase synchronization on the Fourier plane optimizes the PSF. The target profile of the phase change includes two parts, namely, the part to correct the phase nonuniformity in an ideal high-NA system and the part to correct the optical aberration of the imaging system. While the former part can be calculated \cite{He2021JPD} (FIG. S1 \cite{SM}), the latter part requires delicate experimental characterization. We use Zernike polynomials $z_n^m(\rho,\theta)$ to decompose the aberration \cite{Lakshminarayanan2011JMO} and experimentally determine the coefficients of each order $(n, m)$ for the terms up to $n$ = 6 as described in Supplementary Section S6 \cite{SM}. Our study shows that the primary astigmatism ($z_2^2$ and $z_2^{-2}$) and spherical aberration ($z_4^0$) are the major terms in our imaging system (FIG. S12 \cite{SM}). Their corresponding coefficients are determined to be 0.34$\pi$, -0.15$\pi$, and -0.41$\pi$, respectively. The resulting target profile of phase change is displayed in FIG. \ref{FIG2}(b) (with Zernike polynomials up to $n$ = 6).  FIG. \ref{FIG3}(g)$\sim$\ref{FIG3}(k) show the iSCAT contrast images after phase synchronization for the same 40 nm GNP at defocus positions of $\Delta z$ = -1.0 $\mu$m, -0.6 $\mu$m, -0.2 $\mu$m, 0.2 $\mu$m, and 0.6 $\mu$m, respectively.  The iSCAT PSF exhibits much better circular symmetry and tighter focus at $\Delta z$ = - 0.2 $\mu$m, which increases the interference contrast from -0.11 in FIG. \ref{FIG3}(c) to -0.18 in FIG. \ref{FIG3}(i), i.e., over 50\% enhancement. Moreover, the improvement of the PSF leads to a better precision in particle localization, reducing the localization uncertainty \cite{Kashkanova2021OE} from (5.1 nm, 1.2 nm) to (1.2 nm, 0.8 nm) in $(x, y)$ directions, as shown in FIG. \ref{FIG3}(f) and \ref{FIG3}(l), respectively.

\begin{figure}
	\centering
	\includegraphics{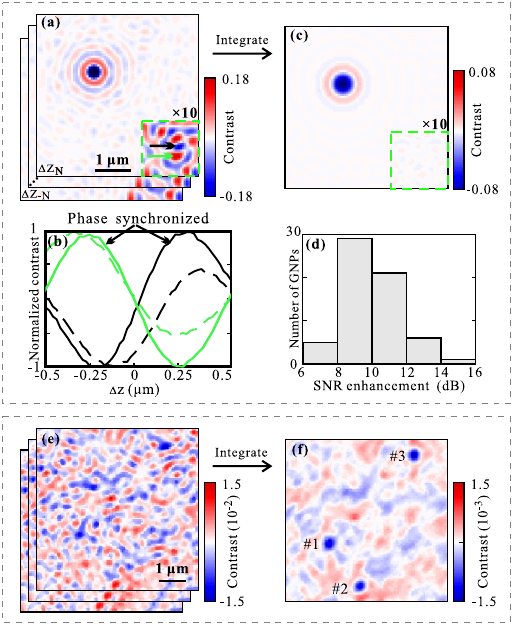}
	\caption{(a) iSCAT contrast images at varied defocus values for a 40 nm GNP on a coverslip after phase synchronization. (b) iSCAT contrast as a function of $\Delta z$ at the positions denoted by the two arrows. Dashed and solid lines denote before and after phase synchronization, respectively. (c) Contrast image obtained by integrating the iSCAT contrast in (a) over $\Delta z$ = -0.5 $\rm{\mu m}$ to 0.5 $\rm{\mu m}$. (d) Histogram of the SBR enhancements by phase synchronization and defocus integration obtained from measuring 61 individual 40 nm GNPs. (e) iSCAT contrast images for an area of a coverslip spin-coated with 10 nm GNPs. (f) Contrast image obtained after integrating the contrast images in (e) over the defocus. The 10 nm GNPs become clearly visible. 
	}
	\label{FIG4}
\end{figure}

We set the synchronized phase difference to $\phi_F=\pi/2$ to utilize the property of anti-symmetric contrast evolution with the defocus. The sub-nanometer surface undulations of a polished substrate can be considered as a collection of point scatterers lying on a perfectly flat substrate surface \cite{Lin2022PRL}, which produce speckles in the contrast image. We thus expect that the contrasts of the speckle background exhibit the anti-symmetric property as shown by the red trace in FIG. \ref{FIG1}(k). We have recorded 51 iSCAT contrast images as the defocus $\Delta z$ changes from -0.5 $\mu$m to 0.5 $\mu$m at a step of 20 nm as illustrated in FIG. \ref{FIG4}(a). The inset at the right bottom corner of FIG. \ref{FIG4}(a) displays the speckle background (enlarged by 10 times for clarity). The color-coded solid lines in FIG. \ref{FIG4}(b) plot the contrast as a function of $\Delta z$ for two points indicated by the arrows in the speckle region. The dashed lines are the plots of the iSCAT contrast with no phase synchronization. Clearly, the solid traces exhibit an anti-symmetric dependence on the defocus. The scattering phase of the 40 nm GNPs at the BFP with $\Delta z$ = 0 corresponds to 0.87$\pi$ after the synchronization due to the complex permittivity of gold ($\epsilon_G=-2.1+4.1i$ \cite{Johnson1972PRB}) and their size (FIG. S2 \cite{SM}). The contrasts from the GNPs do not have the anti-symmetric dependence on the defocus. Thus, an integration of the contrast over the defocus will substantially diminish the speckle background but retain the signals from the GNPs. Consequently, enhancement of signal-to-background ratio (SBR) could be achieved. FIG 4(c) shows the contrast image obtained after the integration, showing successful suppression of the speckles. To quantify the highly reproducible and effective speckle suppression, we measured 61 GNPs (FIG. S16 \cite{SM}), and FIG. \ref{FIG4}(d) presents the statistics of SBR enhancements, yielding a mean enhancement of 10.1 dB. 

The effective suppression of the speckle background enables the detection of substrate-adhered small nanoparticles that are previously indecipherable in the speckle background. We performed measurements on 10 nm GNPs immobilized on a glass coverslip via spin coating. The TEM image of the GNPs (FIG. S3 \cite{SM}) shows that the GNPs have an average size of 9.6 nm with a standard deviation of 0.9 nm. FIG. \ref{FIG4}(e) shows the iSCAT contrast images of a sample on the coverslip, where the signals from the GNPs are not discernible from the speckle background. In fact, our experiments and calculations show that nanoparticles smaller than 17 nm are hardly discernible from the speckle background (Supplementary Section S8.4 \cite{SM}). FIG. \ref{FIG4}(f) displays the contrast image after the integration over the defocus, which clearly shows three 10 nm GNPs on the surface with SBRs over 9 dB.

We have introduced and experimentally demonstrated the concept of Fourier-plane phase synchronization for interferometric scattering microscopy. By combining a photothermal phase plate with a scanning heating laser, we achieve on-demand phase synchronization in practical high–NA microscopy systems subject to aberrations. The realized phase synchronization substantially improves the PSF, yielding near-perfect circular symmetry and over 50\% enhancement in interference contrast, which in turn enables more precise localization of single nanoparticles. Moreover, by locking the phase difference to $\pi/2$, we effectively suppress speckle backgrounds through defocus integration, achieving an SBR improvement exceeding 10 dB and allowing the detection of 10 nm immobilized particles on glass substrates. Beyond this technical advancement in iSCAT, our method provides a conceptually transferable framework for real-time phase control on the Fourier plane in optical microscopy. The integration of photothermal modulation with Fourier-plane phase engineering could further benefit super-resolution imaging, adaptive optics, and wavefront shaping in complex media. We anticipate that the concept and technique of dynamic seamless phase synchronization will serve as a valuable asset for a broad class of optical microscopy modalities.

We acknowledge financial support from the National Natural Science Foundation of China (62235006, 62405106, 12525414, 12011530395), the China Postdoctoral Science Foundation (2024T170295, GZC20230887), State Key Laboratory of Precision Measurement Technology and Instruments (2024PMTI05), the Fundamental Research Funds for the Central Universities (2023BR003, 2024BRB002) through Huazhong University of Science and Technology. The Ministry of Education, Youth and Sports (Project ERC-CZ LL2409), Czech Science Foundation (Project No. 23-07703S) and the Czech Academy of Sciences under the bilateral collaboration project No. NSFC-21-18. We thank Dr. Milan Vala for useful discussions. 

\bibliographystyle{apsrev4-2}
\bibliography{Jiao02_1010_Abb.bib}

@article{Avci2017Optica,
   author = {Avci, Oguzhan and Campana, Maria I. and Yurdakul, Celalettin and Selim \"{U}nl\"{u}, M.},
   title = {Pupil function engineering for enhanced nanoparticle visibility in wide-field interferometric microscopy},
   journal = {Optica},
   volume = {4},
   number = {2},
   pages = {6094-6114},
   ISSN = {2334-2536},
   DOI = {10.1364/optica.4.000247},
   year = {2017},
   type = {Journal Article}
}

@article{Brandt1973AO,
   author = {Brandt, G. B. and Supertzi, E. P. and Henningsen, T.},
   title = {Substrate Cleaning for Integrated Optical Waveguides},
   journal = {Appl. Opt.},
   volume = {12},
   number = {12},
   pages = {2898-2900},
   DOI = {10.1364/AO.12.002898},
   url = {https://opg.optica.org/ao/abstract.cfm?URI=ao-12-12-2898},
   year = {1973},
   type = {Journal Article}
}

@article{Bujak2021SmallMethods,
   author = {Bujak, {\L}ukasz and Holanov{\'a}, Krist{\'y}na and Garc{\'i}a Mar{\'i}n, Antonio and Henrichs, Verena and Barv{\'i}k, Ivan and Braun, Marcus and L{\'a}nsk{\'y}, Zden{\u{e}}k and Piliarik, Marek},
   title = {Fast Leaps between Millisecond Confinements Govern Ase1 Diffusion along Microtubules},
   journal = {Small Methods},
   volume = {5},
   number = {10},
   pages = {2100370},
   ISSN = {2366-9608},
   DOI = {https://doi.org/10.1002/smtd.202100370},
   url = {https://doi.org/10.1002/smtd.202100370},
   year = {2021},
   type = {Journal Article}
}

@article{Cho2018OL,
   author = {Cho, S. and Lee, J. and Kim, H. and Lee, S. and Ohkubo, A. and Lee, J. and Kim, T. and Bae, S. and Joo, W.},
   title = {Super-contrast-enhanced darkfield imaging of nano objects through null ellipsometry},
   journal = {Opt. Lett.},
   volume = {43},
   number = {23},
   pages = {5701-5704},
   ISSN = {1539-4794 (Electronic)
0146-9592 (Linking)},
   DOI = {10.1364/OL.43.005701},
   url = {https://www.ncbi.nlm.nih.gov/pubmed/30499972},
   year = {2018},
   type = {Journal Article}
}

@article{Cole2011NatProt,
   author = {Cole, Richard W. and Jinadasa, Tushare and Brown, Claire M.},
   title = {Measuring and interpreting point spread functions to determine confocal microscope resolution and ensure quality control},
   journal = {Nat. Protoc.},
   volume = {6},
   number = {12},
   pages = {1929-1941},
   ISSN = {1750-2799},
   DOI = {10.1038/nprot.2011.407},
   url = {https://doi.org/10.1038/nprot.2011.407},
   year = {2011},
   type = {Journal Article}
}

@article{Zhou2019APLPhotonics,
   author = {Zhou, Yongzhuang and Handley, Michael and Carles, Guillem and Harvey, Andrew R.},
   title = {Advances in 3D single particle localization microscopy},
   journal = {APL Photonics},
   volume = {4},
   number = {6},
   pages = {060901},
   ISSN = {2378-0967},
   DOI = {10.1063/1.5093310},
   url = {https://doi.org/10.1063/1.5093310},
   year = {2019},
   type = {Journal Article}
}

@article{Delor2020NatMater,
   author = {Delor, M. and Weaver, H. L. and Yu, Q. and Ginsberg, N. S.},
   title = {Imaging material functionality through three-dimensional nanoscale tracking of energy flow},
   journal = {Nat. Mater.},
   volume = {19},
   number = {1},
   pages = {56-62},
   ISSN = {1476-1122 (Print)
1476-1122 (Linking)},
   DOI = {10.1038/s41563-019-0498-x},
   url = {https://www.ncbi.nlm.nih.gov/pubmed/31591529},
   year = {2020},
   type = {Journal Article}
}

@article{Ginsberg2025NatRevMethodsPrim,
   author = {Ginsberg, Naomi S. and Hsieh, Chia-Lung and Kukura, Philipp and Piliarik, Marek and Sandoghdar, Vahid},
   title = {Interferometric scattering microscopy},
   journal = {Nat. Rev. Methods Primers},
   volume = {5},
   number = {1},
   pages = {23},
   ISSN = {2662-8449},
   DOI = {10.1038/s43586-025-00391-1},
   url = {https://doi.org/10.1038/s43586-025-00391-1},
   year = {2025},
   type = {Journal Article}
}

@article{Gruber2024Nature,
   author = {Gruber, Christoph G. and Frey, Laura and Guntermann, Roman and Medina, Dana D. and Cortés, Emiliano},
   title = {Early stages of covalent organic framework formation imaged in operando},
   journal = {Nature},
   volume = {630},
   number = {8018},
   pages = {872-877},
   ISSN = {1476-4687},
   DOI = {10.1038/s41586-024-07483-0},
   url = {https://doi.org/10.1038/s41586-024-07483-0},
   year = {2024},
   type = {Journal Article}
}

@article{Hampson2021NatRevMethodsPrim,
   author = {Hampson, Karen M. and Turcotte, Raphaël and Miller, Donald T. and Kurokawa, Kazuhiro and Males, Jared R. and Ji, Na and Booth, Martin J.},
   title = {Adaptive optics for high-resolution imaging},
   journal = {Nat. Rev. Methods Primers},
   volume = {1},
   number = {1},
   pages = {68},
   ISSN = {2662-8449},
   DOI = {10.1038/s43586-021-00066-7},
   url = {https://doi.org/10.1038/s43586-021-00066-7},
   year = {2021},
   type = {Journal Article}
}

@article{He2021JPD,
   author = {He, Yong and Lin, Shupei and Marc Louis Robert, Hadrien and Li, Hong and Zhang, Pu and Piliarik, Marek and Chen, Xue-Wen},
   title = {Multiscale modeling and analysis for high-fidelity interferometric scattering microscopy},
   journal = {J. Phys. D: Appl. Phys},
   volume = {54},
   number = {27},
   pages = {274002},
   ISSN = {0022-3727
1361-6463},
   DOI = {10.1088/1361-6463/abf70d},
   url = {https://iopscience.iop.org/article/10.1088/1361-6463/abf70d},
   year = {2021},
   type = {Journal Article}
}

@article{Jiao2024OL,
   author = {Jiao, Nanfang and Lin, Shupei and Feng, Delong and He, Yong and Chen, Xue-Wen},
   title = {Defocus-integration interferometric scattering microscopy for speckle suppression and enhancing nanoparticle detection on a substrate},
   journal = {Opt. Lett.},
   volume = {49},
   number = {10},
   pages = {2841-2844},
   DOI = {10.1364/OL.519263},
   url = {https://opg.optica.org/ol/abstract.cfm?URI=ol-49-10-2841},
   year = {2024},
   type = {Journal Article}
}

@article{Johnson1972PRB,
   author = {Johnson, P. B. and Christy, R. W.},
   title = {Optical Constants of the Noble Metals},
   journal = {Phys. Rev. B},
   volume = {6},
   number = {12},
   pages = {4370-4379},
   DOI = {10.1103/PhysRevB.6.4370},
   url = {https://link.aps.org/doi/10.1103/PhysRevB.6.4370},
   year = {1972},
   type = {Journal Article}
}

@article{Kashkanova2021OE,
   author = {Kashkanova, Anna D. and Shkarin, Alexey B. and Mahmoodabadi, Reza Gholami and Blessing, Martin and Tuna, Yazgan and Gemeinhardt, André and Sandoghdar, Vahid},
   title = {Precision single-particle localization using radial variance transform},
   journal = {Opt. Express},
   volume = {29},
   number = {7},
   pages = {11070-11083},
   DOI = {10.1364/OE.420670},
   url = {https://opg.optica.org/oe/abstract.cfm?URI=oe-29-7-11070},
   year = {2021},
   type = {Journal Article}
}

@article{KIRSHNER2013JMicroscopy,
   author = {Kirshner, H. and Aguet, F. and Sage, D. and Unser, M.},
   title = {3-D PSF fitting for fluorescence microscopy: implementation and localization application},
   journal = {J. Microsc.},
   volume = {249},
   number = {1},
   pages = {13-25},
   ISSN = {0022-2720},
   DOI = {https://doi.org/10.1111/j.1365-2818.2012.03675.x},
   url = {https://onlinelibrary.wiley.com/doi/abs/10.1111/j.1365-2818.2012.03675.x},
   year = {2013},
   type = {Journal Article}
}

@article{Li2025NatPhotonics,
   author = {Li, Qing-Yue and Lyu, Pin-Tian and Kang, Bin and Chen, Hong-Yuan and Xu, Jing-Juan},
   title = {Electrochemically modulated interferometric scattering microscopy for imaging ion channel activity in live cells},
   journal = {Nat. Photonics},
   volume = {19},
   number = {8},
   pages = {871-878},
   ISSN = {1749-4893},
   DOI = {10.1038/s41566-025-01696-z},
   url = {https://doi.org/10.1038/s41566-025-01696-z},
   year = {2025},
   type = {Journal Article}
}

@article{Lin2022PRL,
   author = {Lin, Shupei and He, Yong and Feng, Delong and Piliarik, Marek and Chen, Xue-Wen},
   title = {Optical Fingerprint of Flat Substrate Surface and Marker-Free Lateral Displacement Detection with Angstrom-Level Precision},
   journal = {Phys. Rev. Lett.},
   volume = {129},
   number = {21},
   pages = {213201},
   DOI = {10.1103/PhysRevLett.129.213201},
   url = {https://link.aps.org/doi/10.1103/PhysRevLett.129.213201},
   year = {2022},
   type = {Journal Article}
}

@article{Meng2021AcsPhotonics,
   author = {Meng, Xuanhui and Sonn-Segev, Adar and Schumacher, Anne and Cole, Daniel and Young, Gavin and Thorpe, Stephen and Style, Robert W. and Dufresne, Eric R. and Kukura, Philipp},
   title = {Micromirror Total Internal Reflection Microscopy for High-Performance Single Particle Tracking at Interfaces},
   journal = {ACS Photonics},
   volume = {8},
   number = {10},
   pages = {3111-3118},
   DOI = {10.1021/acsphotonics.1c01268},
   url = {https://doi.org/10.1021/acsphotonics.1c01268},
   year = {2021},
   type = {Journal Article}
}

@article{Merryweather2021Nature,
   author = {Merryweather, A. J. and Schnedermann, C. and Jacquet, Q. and Grey, C. P. and Rao, A.},
   title = {Operando optical tracking of single-particle ion dynamics in batteries},
   journal = {Nature},
   volume = {594},
   number = {7864},
   pages = {522-528},
   ISSN = {1476-4687 (Electronic)
0028-0836 (Linking)},
   DOI = {10.1038/s41586-021-03584-2},
   url = {https://www.ncbi.nlm.nih.gov/pubmed/34163058},
   year = {2021},
   type = {Journal Article}
}

@book{Mertz2019IntrodOptMicroscopy,
   author = {Mertz, Jerome},
   title = {Introduction to Optical Microscopy},
   publisher = {Cambridge University Press},
   address = {Cambridge},
   edition = {2},
   ISBN = {9781108428309},
   DOI = {DOI: 10.1017/9781108552660},
   url = {https://www.cambridge.org/core/product/F6C6318C87732519D7E07BA7A03F0B81},
   year = {2019},
   type = {Book}
}

@article{Park2018NatPhotonics,
   author = {Park, YongKeun and Depeursinge, Christian and Popescu, Gabriel},
   title = {Quantitative phase imaging in biomedicine},
   journal = {Nat. Photonics},
   volume = {12},
   number = {10},
   pages = {578-589},
   ISSN = {1749-4893},
   DOI = {10.1038/s41566-018-0253-x},
   url = {https://doi.org/10.1038/s41566-018-0253-x},
   year = {2018},
   type = {Journal Article}
}

@article{Piliarik2014NatComm,
   author = {Piliarik, Marek and Sandoghdar, Vahid},
   title = {Direct optical sensing of single unlabelled proteins and super-resolution imaging of their binding sites},
   journal = {Nat. Commun.},
   volume = {5},
   number = {1},
   pages = {4495},
   ISSN = {2041-1723},
   DOI = {10.1038/ncomms5495},
   url = {https://doi.org/10.1038/ncomms5495},
   year = {2014},
   type = {Journal Article}
}

@article{Ortega2014NanoLett,
   author = {Ortega Arroyo, J. and Andrecka, J. and Spillane, K. M. and Billington, N. and Takagi, Y. and Sellers, J. R. and Kukura, P.},
   title = {Label-free, all-optical detection, imaging, and tracking of a single protein},
   journal = {Nano Lett.},
   volume = {14},
   number = {4},
   pages = {2065-70},
   ISSN = {1530-6992 (Electronic)
1530-6984 (Linking)},
   DOI = {10.1021/nl500234t},
   url = {https://www.ncbi.nlm.nih.gov/pubmed/24597479},
   year = {2014},
   type = {Journal Article}
}

@book{Novotny2012PrinNanoOpt,
   author = {Novotny, L. and Hecht, B. },
   title = {Principles of nano-optics},
   publisher = {University Press, Cambridge},
   address = {United Kingdom},
   edition = {2nd edition},
   year = {2012},
   url = {https://doi.org/10.1017/CBO9780511794193},
   type = {Book}
}

@book{Reinhardt2018HBSiWaferCleanTech,
   author = {Reinhardt, Karen and Kern, Werner},
   title = {Handbook of silicon wafer cleaning technology},
   publisher = {William Andrew},
   ISBN = {032351085X},
   year = {2018},
   type = {Book},
   url = {https://www.sciencedirect.com/book/9780323510844/handbook-of-silicon-wafer-cleaning-technology}
}

@article{Robert2022AOM,
   author = {Robert, Hadrien M. L. and {\u{C}}i{\u{c}}ala, Martin and Piliarik, Marek},
   title = {Shaping of Optical Wavefronts Using Light-Patterned Photothermal Metamaterial},
   journal = {Adv. Opt. Mater.},
   volume = {10},
   number = {21},
   pages = {2200960},
   ISSN = {2195-1071},
   DOI = {https://doi.org/10.1002/adom.202200960},
   url = {https://doi.org/10.1002/adom.202200960},
   year = {2022},
   type = {Journal Article}
}

@article{Sofronov2021PRAppl,
   author = {Sofronov, Anton and Afinogenov, Boris and Medvedev, Anton and Shorokhov, Aleksandr and Riabko, Maksim and Polonsky, Stanislav},
   title = {Optical Detection of Deeply Subwavelength Nanoparticles for Silicon Metrology},
   journal = {Phys. Rev. Appl.},
   volume = {15},
   number = {6},
   pages = {064049},
   DOI = {10.1103/PhysRevApplied.15.064049},
   url = {https://link.aps.org/doi/10.1103/PhysRevApplied.15.064049},
   year = {2021},
   type = {Journal Article}
}

@inbook{Su2022EquipMaterClean,
   author = {Su, Yikai and Zhang, Yong and Su, Yikai and Zhang, Yong},
   title = {Equipment and Materials in Cleanroom},
   booktitle = {Passive Silicon Photonic Devices: Design, Fabrication, and Testing},
   publisher = {AIP Publishing LLC},
   pages = {99-124},
   ISBN = {978-0-7354-2428-9},
   DOI = {10.1063/9780735424319_005},
   url = {https://doi.org/10.1063/9780735424319_005},
   year = {2022},
   type = {Book Section}
}

@article{Takeda1982JOSA,
   author = {Takeda, Mitsuo and Ina, Hideki and Kobayashi, Seiji},
   title = {Fourier-transform method of fringe-pattern analysis for computer-based topography and interferometry},
   journal = {J. Opt. Soc. Am.},
   volume = {72},
   number = {1},
   pages = {156-160},
   DOI = {10.1364/JOSA.72.000156},
   url = {https://opg.optica.org/abstract.cfm?URI=josa-72-1-156},
   year = {1982},
   type = {Journal Article}
}

@article{Taylor2019NatPhotonics,
   author = {Taylor, Richard W. and Mahmoodabadi, Reza Gholami and Rauschenberger, Verena and Giessl, Andreas and Schambony, Alexandra and Sandoghdar, Vahid},
   title = {Interferometric scattering microscopy reveals microsecond nanoscopic protein motion on a live cell membrane},
   journal = {Nat. Photonics},
   volume = {13},
   number = {7},
   pages = {480-487},
   ISSN = {1749-4885
1749-4893},
   DOI = {10.1038/s41566-019-0414-6},
   url = {https://doi.org/10.1038/s41566-019-0414-6},
   year = {2019},
   type = {Journal Article}
}

@article{Yang2023OptElectScience,
   author = {Yang, Yiqian and Forbes, Andrew and Cao, Liangcai},
   title = {A review of liquid crystal spatial light modulators: devices and applications},
   journal = {Opto-Electron. Sci},
   volume = {2},
   number = {8},
   pages = {230026},
   ISSN = {2097-0382},
   DOI = {10.29026/oes.2023.230026},
   url = {https://www.oejournal.org/article/doi/10.29026/oes.2023.230026},
   year = {2023},
   type = {Journal Article}
}

@article{Young2018Science,
   author = {Young, Gavin and Hundt, Nikolas and Cole, Daniel and Fineberg, Adam and Andrecka, Joanna and Tyler, Andrew and Olerinyova, Anna and Ansari, Ayla and Marklund, Erik G. and Collier, Miranda P. and Chandler, Shane A. and Tkachenko, Olga and Allen, Joel and Crispin, Max and Billington, Neil and Takagi, Yasuharu and Sellers, James R. and Eichmann, Cédric and Selenko, Philipp and Frey, Lukas and Riek, Roland and Galpin, Martin R. and Struwe, Weston B. and Benesch, Justin L. P. and Kukura, Philipp},
   title = {Quantitative mass imaging of single biological macromolecules},
   journal = {Science},
   volume = {360},
   number = {6387},
   pages = {423-427},
   DOI = {10.1126/science.aar5839 %J Science},
   url = {https://science.sciencemag.org/content/sci/360/6387/423.full.pdf},
   year = {2018},
   type = {Journal Article}
}

@article{Zhu2022IJEM,
   author = {Zhu, Jinlong and Liu, Jiamin and Xu, Tianlai and Yuan, Shuai and Zhang, Zexu and Jiang, Hao and Gu, Honggang and Zhou, Renjie and Liu, Shiyuan},
   title = {Optical wafer defect inspection at the 10 nm technology node and beyond},
   journal = {Int. J. Extreme Manuf.},
   volume = {4},
   number = {3},
   pages = {032001},
   ISSN = {2631-8644
2631-7990},
   DOI = {10.1088/2631-7990/ac64d7},
   url = {http://dx.doi.org/10.1088/2631-7990/ac64d7},
   year = {2022},
   type = {Journal Article}
}

@article{Barr2009CurrentProtCellBio,
   author = {Barr, Valarie A. and Bunnell, Stephen C.},
   title = {Interference Reflection Microscopy},
   journal = {Curr. Protoc. Cell Biol.},
   volume = {45},
   number = {1},
   pages = {4.23.1-4.23.19},
   ISSN = {1934-2500},
   DOI = {https://doi.org/10.1002/0471143030.cb0423s45},
   url = {https://doi.org/10.1002/0471143030.cb0423s45},
   year = {2009},
   type = {Journal Article}
}

@inbook{Kim2021DigHoloMicro,
   author = {Kim, Myung K.},
   title = {Digital Holographic Microscopy},
   booktitle = {Digital Holographic Microscopy: Principles, Techniques, and Applications},
   editor = {Kim, Myung K.},
   publisher = {Springer New York},
   address = {New York, NY},
   pages = {149-190},
   ISBN = {978-1-4419-7793-9},
   DOI = {10.1007/978-1-4419-7793-9_11},
   url = {https://doi.org/10.1007/978-1-4419-7793-9_11},
   year = {2011},
   type = {Book Section}
}

@article{Kashkanova2022NatMethod,
   author = {Kashkanova, Anna D. and Blessing, Martin and Gemeinhardt, André and Soulat, Didier and Sandoghdar, Vahid},
   title = {Precision size and refractive index analysis of weakly scattering nanoparticles in polydispersions},
   journal = {Nat. Methods},
   volume = {19},
   number = {5},
   pages = {586-593},
   ISSN = {1548-7091
1548-7105},
   DOI = {10.1038/s41592-022-01460-z},
   year = {2022},
   type = {Journal Article}
}

@article{Huang2024LSA,
   author = {Huang, Mingchuan and Chen, Qiankun and Liu, Yang and Zhang, Chi and Zhang, Rongjin and Yuan, Junhua and Zhang, Douguo},
   title = {One-dimensional photonic crystal enhancing spin-to-orbital angular momentum conversion for single-particle tracking},
   journal = {Light:Sci. Appl.},
   volume = {13},
   number = {1},
   pages = {268},
   ISSN = {2047-7538},
   DOI = {10.1038/s41377-024-01623-x},
   url = {https://doi.org/10.1038/s41377-024-01623-x},
   year = {2024},
   type = {Journal Article}
}

@article{Huang2017ACSNano,
   author = {Huang, Y. F. and Zhuo, G. Y. and Chou, C. Y. and Lin, C. H. and Chang, W. and Hsieh, C. L.},
   title = {Coherent Brightfield Microscopy Provides the Spatiotemporal Resolution To Study Early Stage Viral Infection in Live Cells},
   journal = {ACS Nano},
   volume = {11},
   number = {3},
   pages = {2575-2585},
   
   ISSN = {1936-086X (Electronic)
    1936-0851 (Linking)},
   DOI = {10.1021/acsnano.6b05601},
   url = {https://www.ncbi.nlm.nih.gov/pubmed/28067508},
   year = {2017},
   type = {Journal Article}
}

@article{Eismann2021LSA,
   author = {Eismann, J\"{o}rg S. and Neugebauer, Martin and Mantel, Klaus and Banzer, Peter},
   title = {Absolute characterization of high numerical aperture microscope objectives utilizing a dipole scatterer},
   journal = {Light:Sci. Appl.},
   volume = {10},
   number = {1},
   pages = {223},
   ISSN = {2047-7538},
   DOI = {10.1038/s41377-021-00663-x},
   url = {https://doi.org/10.1038/s41377-021-00663-x},
   year = {2021},
   type = {Journal Article}
}

@article{Shechtman2014PRL,
  title = {Optimal Point Spread Function Design for 3D Imaging},
  author = {Shechtman, Yoav and Sahl, Steffen J. and Backer, Adam S. and Moerner, W. E.},
  journal = {Phys. Rev. Lett.},
  volume = {113},
  issue = {13},
  pages = {133902},
  numpages = {5},
  year = {2014},
  month = {Sep},
  publisher = {American Physical Society},
  doi = {10.1103/PhysRevLett.113.133902},
  url = {https://link.aps.org/doi/10.1103/PhysRevLett.113.133902}
}

@article{Xie2014PRL,
  title = {Harnessing the Point-Spread Function for High-Resolution Far-Field Optical Microscopy},
  author = {Xie, Xiangsheng and Chen, Yongzhu and Yang, Ken and Zhou, Jianying},
  journal = {Phys. Rev. Lett.},
  volume = {113},
  issue = {26},
  pages = {263901},
  numpages = {5},
  year = {2014},
  month = {Dec},
  publisher = {American Physical Society},
  doi = {10.1103/PhysRevLett.113.263901},
  url = {https://link.aps.org/doi/10.1103/PhysRevLett.113.263901}
}

@article{Parthasarathy2012NatMethod,
   author = {Parthasarathy, Raghuveer},
   title = {Rapid, accurate particle tracking by calculation of radial symmetry centers},
   journal = {Nat. Methods},
   volume = {9},
   number = {7},
   pages = {724-726},
   ISSN = {1548-7105},
   DOI = {10.1038/nmeth.2071},
   url = {https://doi.org/10.1038/nmeth.2071},
   year = {2012},
   type = {Journal Article}
}

@article{Diezmann2017ChemRev,
   author = {von Diezmann, Lexy and Shechtman, Yoav and Moerner, W. E.},
   title = {Three-Dimensional Localization of Single Molecules for Super-Resolution Imaging and Single-Particle Tracking},
   journal = {Chem. Rev.},
   volume = {117},
   number = {11},
   pages = {7244-7275},
   ISSN = {0009-2665},
   DOI = {10.1021/acs.chemrev.6b00629},
   url = {https://doi.org/10.1021/acs.chemrev.6b00629},
   year = {2017},
   type = {Journal Article}
}

@article{Lakshminarayanan2011JMO,
   author = {Lakshminarayanan, Vasudevan and Fleck, Andre},
   title = {Zernike polynomials: a guide},
   journal = {J. Mod. Opt.},
   volume = {58},
   number = {7},
   pages = {545-561},
   ISSN = {0950-0340},
   DOI = {10.1080/09500340.2011.554896},
   url = {https://doi.org/10.1080/09500340.2011.554896},
   year = {2011},
   type = {Journal Article}
}

@article{Failla2006NanoLett,
   author = {Failla, Antonio Virgilio and Qian, Hui and Qian, Huihong and Hartschuh, Achim and Meixner, Alfred J.},
   title = {Orientational Imaging of Subwavelength Au Particles with Higher Order Laser Modes},
   journal = {Nano Lett.},
   volume = {6},
   number = {7},
   pages = {1374-1378},
   ISSN = {1530-6984},
   DOI = {10.1021/nl0603404},
   url = {https://doi.org/10.1021/nl0603404},
   year = {2006},
   type = {Journal Article}
}

@article{Wang2021AnalChem,
   author = {Wang, X. and Zeng, Q. and Xie, F. and Wang, J. and Yang, Y. and Xu, Y. and Li, J. and Yu, H.},
   title = {Automated Nanoparticle Analysis in Surface Plasmon Resonance Microscopy},
   journal = {Anal. Chem.},
   volume = {93},
   number = {20},
   pages = {7399-7404},
  ISSN = {1520-6882 (Electronic)
               0003-2700 (Linking)},
   DOI = {10.1021/acs.analchem.1c01493},
   url = {https://www.ncbi.nlm.nih.gov/pubmed/33973472},
   year = {2021},
   type = {Journal Article}
}

@article{Weisenburger2015ContempPhys,
   author = {Weisenburger, Siegfried and Sandoghdar, Vahid},
   title = {Light microscopy: an ongoing contemporary revolution},
   journal = {Contemp. Phys.},
   volume = {56},
   number = {2},
   pages = {123-143},
   ISSN = {0010-7514
1366-5812},
   DOI = {10.1080/00107514.2015.1026557},
   year = {2015},
   type = {Journal Article}
}

@article{Gjonaj2013PRL,
  title = {Focusing and Scanning Microscopy with Propagating Surface Plasmons},
  author = {Gjonaj, B. and Aulbach, J. and Johnson, P. M. and Mosk, A. P. and Kuipers, L. and Lagendijk, A.},
  journal = {Phys. Rev. Lett.},
  volume = {110},
  issue = {26},
  pages = {266804},
  numpages = {5},
  year = {2013},
  month = {Jun},
  publisher = {American Physical Society},
  doi = {10.1103/PhysRevLett.110.266804},
  url = {https://link.aps.org/doi/10.1103/PhysRevLett.110.266804}
}

@article{Yu2010PRL,
  title = {Direct Subangstrom Measurement of Surfaces of Oxide Particles},
  author = {Yu, R. and Hu, L. H. and Cheng, Z. Y. and Li, Y. D. and Ye, H. Q. and Zhu, J.},
  journal = {Phys. Rev. Lett.},
  volume = {105},
  issue = {22},
  pages = {226101},
  numpages = {4},
  year = {2010},
  month = {Nov},
  publisher = {American Physical Society},
  doi = {10.1103/PhysRevLett.105.226101},
  url = {https://link.aps.org/doi/10.1103/PhysRevLett.105.226101}
}

@article{Pushkina2021PRL,
  title = {Superresolution Linear Optical Imaging in the Far Field},
  author = {Pushkina, A. A. and Maltese, G. and Costa-Filho, J. I. and Patel, P. and Lvovsky, A. I.},
  journal = {Phys. Rev. Lett.},
  volume = {127},
  issue = {25},
  pages = {253602},
  numpages = {6},
  year = {2021},
  month = {Dec},
  publisher = {American Physical Society},
  doi = {10.1103/PhysRevLett.127.253602},
  url = {https://link.aps.org/doi/10.1103/PhysRevLett.127.253602}
}

@article{Roggemann1997RevModPhys,
  title = {Improving the resolution of ground-based telescopes},
  author = {Roggemann, Michael C. and Welsh, Byron M. and Fugate, Robert Q.},
  journal = {Rev. Mod. Phys.},
  volume = {69},
  issue = {2},
  pages = {437--506},
  numpages = {0},
  year = {1997},
  month = {Apr},
  publisher = {American Physical Society},
  doi = {10.1103/RevModPhys.69.437},
  url = {https://link.aps.org/doi/10.1103/RevModPhys.69.437}
}

@msi{SM,
  title={See Supplementary Materials for Photothermal Fourier-plane Phase Synchronization for Interferometric Scattering Microscopy},
  author={Lin,S and others}
}

\end{document}